\documentstyle[12pt]{article}

\begin{document}

\setcounter{page}{1}

\pagestyle{plain} \vspace{1cm}
\begin{center}
\Large{\bf Nonminimal Inflation on the Randall-Sundrum II Brane with Induced Gravity}\\
\small \vspace{1cm} {\bf Kourosh Nozari $^{a,b}$ }\quad and \quad {\bf M. Shoukrani$^{a}$}\\
\vspace{0.5cm} {\it $^{a}$Department of Physics, Faculty of Basic
Sciences,\\
University of Mazandaran,\\
P. O. Box 47416-95447, Babolsar, IRAN}\\
\vspace{0.3cm} {\it $^{b}$Research Institute for Astronomy and
Astrophysics of
Maragha, \\P. O. Box 55134-441, Maragha, IRAN\\
 knozari@umz.ac.ir}
\end{center}
\vspace{1.5cm}
\begin{abstract}
We study an inflation model that inflaton field is non-minimally
coupled to the induced scalar curvature on the Randall-Sundrum (RS)
II brane. We investigate the effects of the non-minimal coupling on
the inflationary dynamics of this braneworld model. Our study shows
that the number of e-folds decreases by increasing the value of the
non-minimal coupling. We compare our model parameters with the
minimal case and also with recent observational data. In comparison
with recent observation, we obtain a constraint on the values that
the non-minimal coupling attains.
\end{abstract}
\newpage
\section{Introduction}
It is well known that inflation provides a natural explanation for
some long-standing problems of the standard hot big bang cosmology.
Inflation is also a successful scenario for production and evolution
of the perturbations in primary stages of the universe evolution
[1]. Although inflation paradigm is successful in these respects,
there is a problem for realization of this scenario that we do not
know how to integrate it with ideas of the particle physics [2,3,4].
One important feature of the inflationary paradigm is the fact that
inflaton can interact with other fields such as gravitational sector
of the theory. This interaction is shown by the non-minimal coupling
of the inflaton field and Ricci scalar in the spirit of the
scalar-tensor theories. In fact, there are several compelling
reasons for inclusion of an explicit non-minimal coupling of the
inflaton field and gravity in the action. For instance, non-minimal
coupling arises at the quantum level when quantum corrections to the
scalar field theory are considered. Even if for the classical,
unperturbed theory this non-minimal coupling vanishes, it is
necessary for the renormalizability of the scalar field theory in
curved space. In most theories used to describe inflationary
scenarios, it turns out that a non-vanishing value of the coupling
constant cannot be avoided. Also, in general relativity, and in all
other metric theories of gravity in which the scalar field is not
part of the gravitational sector, the coupling constant necessarily
assumes the value of \, $\frac{1}{6}$\, ( see [5,6] for more
details). Thus, it is natural to study an extension of the inflation
proposal that contains explicit non-minimal coupling of the scalar
field and gravity. Currently, theories of extra spatial dimensions,
in which the observed universe is realized as a brane embedded in a
higher dimensional spacetime, have attracted a lot of attention. In
this framework, ordinary matters are trapped on the brane, but
gravitation propagates through the entire spacetime [7-15]. The
cosmological evolution on the brane is given by an effective
Friedmann equation that incorporates the effects of the bulk in a
non-trivial manner [16,17,18,19]. The chaotic inflation on the RSII
brane has been studied for first time in [20]. Then extension of
this study to warm inflation has been preformed in [21].

With these preliminaries, the goal of the present study is to
investigate a braneworld viewpoint of inflation with an explicit
non-minimal coupling of the scalar field and Ricci curvature on the
brane. We study possible impact of the non-minimal coupling on the
dynamics of this braneworld-inspired inflation. Our setup is based
on the RS II braneworld model and we assume that brane is stable and
inflation dynamics on the brane is independent of the bulk dynamics.
We study modifications of this slow-roll inflation due to
non-minimal coupling of the inflaton field and gravity on the brane.
We study also parameter space of the model numerically to obtain
constraints imposed on this model by recent observations released by
WMAP5+SDSS+SNIa datasets and we compare our results with the minimal
case studied in Ref. [20]. Through this paper a dot on a quantity
marks its time differentiation while a prime denotes differentiation
with respect to the scalar field, $\varphi$.

\section{The Setup}
We start with the following action to construct a braneworld
non-minimal inflation scenario
\begin{equation}
S=\frac{1}{2
\kappa_{5}^{2}}\int_{bulk}d^5X\sqrt{-g^{(5)}}\big(R^{(5)}-2\Lambda_{5}\big)+
\int_{brane}d^4x\sqrt{-g}\bigg[\frac{R}{2\kappa_{4}^2}-\frac{1}{2}g^{\mu\nu}
 \partial_\mu\varphi\partial_\nu\varphi-\frac{1}{2}\xi
 R\varphi^2-V(\varphi)-\lambda\bigg]
\end{equation}
In this action, which is written in the Jordan frame,  $X$ is
coordinate in the bulk, while $x$ shows induced coordinate on the
brane.\, $\kappa_5^2 $ is 5-dimensional gravitational constant,
$R^{(5)}$ is 5-dimensional Ricci scalar and $\xi$ is the nonminimal
coupling (NMC) of the scalar field $\varphi$ and Ricci curvature on
the brane,$\Lambda_{5}$ is the 5-dimensional cosmological constant
in the bulk and $\lambda$ is the brane tension. We have chosen a
conformal coupling of the inflaton and gravity on the brane for
simplicity. In other words, non-minimal coupling of the scalar field
and gravity on the brane is
$\alpha(\varphi)=\frac{1}{2}(1-\xi\varphi^{2})$ with
$\kappa_{4}^{2}=8\pi G=1$. We note that there are two critical
values of $\varphi$ given by $\varphi_{c}=\pm\frac{1}{\sqrt{\xi}}$
that should be avoided to have well-defined field equations.
Variation of the action with respect to $\varphi$ leads to the
equation of dynamics for scalar field on the brane
\begin{equation}
\ddot{\varphi}+3H\dot{\varphi}+\xi R\varphi+\frac{dV}{d\varphi}=0,
\end{equation}
where $R=6(\frac{\ddot{a}}{a}+\frac{\dot{a}^{2}}{a^2})$\, for
spatially flat FRW geometry on the brane and $H=\frac{\dot{a}}{a}$
is the brane Hubble parameter. In the slow-roll approximation where
$\ddot{\varphi}\ll V(\varphi)$, equation of motion for the scalar
field takes the following form
\begin{equation}
\dot{\varphi}=-\frac{\xi R\varphi+V'(\varphi)}{3H}.
\end{equation}
The energy density and pressure of the non-minimally coupled scalar
field are given by (see Ref. [5,6] for discussion on the various
representations of the energy-momentum tensor of a non-minimally
coupled scalar field)
\begin{equation}
\rho_\varphi=\frac{1}{2}\dot{\varphi}^2+V(\varphi)+6\xi
H\varphi\dot{\varphi}+3\xi H^2\varphi^2
\end{equation}
and
\begin{equation}
P_\varphi=\frac{1}{2}\dot{\varphi}^2-V(\varphi)-2\xi
\varphi\ddot{\varphi}-2\xi \dot{\varphi}^2-4\xi
H\varphi\dot{\varphi}-\xi(2\dot{H}+3H^2)\varphi^2
\end{equation}
respectively. The conservation equation for scalar field energy
density in this setup is given by
\begin{equation}
\dot{\rho}_\varphi+3H(\rho_\varphi+P_\varphi)=\dot{\varphi}V'+
\dot{\varphi}\ddot{\varphi}+3H\dot{\varphi}^2+\xi
R\varphi\dot{\varphi}
\end{equation}
where from equation (2) we find
\begin{equation}
\dot{\rho}_\varphi+3H(\rho_\varphi+P_\varphi)=0
\end{equation}
Note that the authors of Ref. [22] have treated this conservation on
the brane in a relatively different way. They have defined a total
energy-momentum tensor which consists of two parts: a pure
(canonical) scalar field energy-momentum tensor and a non-minimal
coupling-dependent part. The total energy density defined in this
manner is then conserved. In our case, we have included all possible
terms in equations (4) and (5) from the beginning and therefore,
total energy density defined in this manner is conserved too.

Now, in a cosmological scenario based on the Randall-Sundrum II
braneworld, the effective Friedmann equation on the brane can be
written as follows [16,17,18,19]
\begin{equation}
H^2=\frac{8\pi}{3M^2}\rho(1+\frac{\rho}{2\lambda}),
\end{equation}
where we have assumed that metric on the brane is spatially flat FRW
type and the effective cosmological constant on the brane is
negligible during inflation phase. $M$ is the four-dimensional Plank
scale, $\rho$ is the total energy density on the brane, namely
$\rho=\rho_\varphi+\rho_m$  where $\rho_{m}$ is energy density of
ordinary matter on the brane and $\lambda$ is 3-brane tension. In
the low energy limit where $\lambda\gg\rho$, one recovers the
standard Friedmann cosmology. In the high energy limit where
$\rho\gg\lambda$, the braneworld effects are dominant. In which
follows we set $\rho_{m}=0$ for simplicity.

In the slow-roll approximation, $\rho_\varphi$ attains the following
approximate form
\begin{equation}
\rho_\varphi\approx V(\varphi)+6\xi H\varphi\dot{\varphi}+3\xi
H^2\varphi^2,
\end{equation}
We use this approximate form in equation (10) to find cosmological
dynamics of the model. Now the Friedmann equation of the model takes
the following form(see[23,24] for more details)
\begin{equation}
H^2+\frac{k}{a^{2}}=\Bigg[r_{c}\alpha(\varphi)\big(H^{2}+\frac{k}{a^{2}}\big)-\frac{\kappa_{5}^2}{6}\big(
 \rho+\lambda \big)\Bigg]^{2}.
\end{equation}
where we assume that the metric on the brane is spatially flat FRW
type $(k=0)$.\,$r_{c}=\frac{\kappa_{5}^{2}}{2\kappa_{4}^{2}}$\,, is
the induced -gravity crossover length scale.
\begin{equation}
    H^2=\Bigg[r_{c}\alpha(\varphi)H^{2}-\frac{\kappa_{5}^2}{6}\bigg(V(\varphi)-2\xi^{2}R\varphi^{2}-2\xi\varphi
 V'+3\xi H^{2}\varphi^{2}+\lambda \bigg)\Bigg]^{2}
\end{equation}
Using the definition of $\dot{\varphi}$ as given by equation (3),
this leads to a forth order equation for Hubble parameter of the
model
$$
H^{4}\Big(r_{c}\alpha(\varphi)-\frac{\xi
\varphi^{2}\kappa_{5}^{2}}{2}\Big)^{2}-H^{2}\Bigg[1+2\Big(r_{c}\alpha(\varphi)-\frac{\xi
\varphi^{2}\kappa_{5}^{2}}{2}\Big)\Bigg(\frac{\kappa_{5}^2}{6}\Big(V(\varphi)-{2\xi^{2}R\varphi^{2}-2\xi\varphi
 V'}+\lambda \Big)\Bigg)\Bigg]
$$

\begin{equation}
+\Bigg[\frac{\kappa_{5}^2}{6}\Big(V(\varphi)-{2\xi^{2}R\varphi^{2}-2\xi\varphi
 V'}+\lambda \Big)\Bigg]^{2}=0
\end{equation}
With a minimally coupled scalar field ($\xi=0$)and in the absence of
induced gravity on the brane (the pure Randall-Sandrum case with
$r_{c}=0$) we find
$$H^{2}\simeq \frac{8\pi}{3M^{2}}
V(\varphi)\Big(1+\frac{V(\varphi)}{2\lambda}\Big),$$
$$\dot{\varphi}\simeq -\frac{V'}{3H}\,\,$$ which are appropriate Friedmann
and scalar field equations for this case [20]. By assuming a
non-vanishing value of the non-minimal coupling $\xi$, we can solve
equation (12) for $H^2$ to find the following the following
solutions
\begin{equation}
H^2=\frac{\lambda}{b^2}\,\Big[\,\frac{8\pi}{m_{p}^2}+\frac{8\pi
r_c}{3 m_{p}^2}
  \,b\,(1+\frac{a}{\lambda})\,\Big]\pm\frac{2}{b}\,\sqrt{\frac{1}{b^2}+\frac{r_c}{b}(1+\frac{a}{\lambda})}
\end{equation}
where $a$ and $b$ are defined as follows
\begin{equation}
a\equiv V-2\xi\varphi(\xi R\varphi+V'),\quad\quad b\equiv 2r_c -
3\xi\varphi^2 \kappa_{4}^{2}.
\end{equation}
To be more specific, in which follows we consider just the plus sign
in equation (13). We note that reality of the solution for $H^{2}$
requires that
$$ \lambda\geq \frac{-\beta a}{1+\beta}$$ where
$\beta\equiv r_{c}b$. Now, we define the slow-roll parameters of our
model as follows
$$\varepsilon\equiv-\frac{\dot{H}}{H^2},$$
$$\eta\equiv-\frac{\ddot{H}}{H\dot{H}}$$
and
$$\gamma^{2}\equiv2\varepsilon\eta-\frac{d\eta}{dt}.$$
We need to calculate these parameters in our non-minimal braneworld
model. For $\varepsilon$ we find
\begin{equation}
\varepsilon=\frac{2\dot{A}B -2\dot{B}A -\dot{C} C^{-1} A}{4H^{3} A},
\end{equation}
where we have defined
$$ A=\frac{\Big(2r_{c}-3\xi \kappa_{4}^{2}\varphi^{2}\Big)^{2}}{\lambda}$$
$$B=\frac{8\pi}{m_{p}^2}+\frac{8\pi r_c}{3 m_{p}^2}\, b\,(1+\frac{a}{\lambda}),$$
and
$$C=\frac{4}{b^4}+\frac{4 r_c}{b^3}\,(1+\frac{a}{\lambda}).$$ The
second slow-roll parameter,\, $\eta$\,, is calculated as follows
$$\eta =\bigg[4A^{2}H^{3}(2\dot{B}A
-2\dot{B}\dot{A}+\dot{C}C^{-1}A)\bigg]^{-1}\bigg[(8 \dot{A} A H^{2}
+2\dot{B} A-2\dot{A} B+\dot{C}C^{-1}A)(2\dot{B}A -2\dot{A}B +\dot{C}
C^{-1}A)
 $$
 \begin{equation}
-(2\ddot{B} A-2\ddot{A}B + \ddot{C} C^{-1} A-\dot{C}^{2}C^{-2}A
+\dot{C} C^{-1} \dot{A}) 4A^{2}H^{2}\bigg].
 \end{equation}
And finally, the third slow-roll parameter, $\gamma^2$, in our model
takes the following form
\begin{equation}
\gamma^2=-\frac{\dot{\varphi}V'''}{3H^{3}}.
\end{equation}
Inflation occurs when the condition  $\varepsilon<1$ ( or
equivalently $|\eta|<1$) is fulfilled. With this condition and using
(15) we find
\begin{equation}
\rho_\varphi< V+3\xi H^{2}\varphi^2.
\end{equation}
This is the condition for realization of the non-minimal inflation
on the brane. Therefore, to have an inflationary period on the
brane, the condition (18) should be fulfilled. Inflationary phase
will terminates when the universe heats up so that the condition
$\varepsilon= 1$ ( or $|\eta|=1$) is satisfied. We should stress
here that in our non-minimal setup, depending on the values of the
non-minimal coupling $\xi$, it is possible to fulfill $|\eta| =1$
condition even before fulfilling $\epsilon=1$. In other words, for
some values of the non-minimal coupling, inflation terminates if the
condition $|\eta|=1$ is fulfilled. Figure $1$ shows the the natural
exit from inflation phase when the condition $|\eta|=1$ is fulfilled
before occurrence of $\epsilon=1$. In this case we have
\begin{equation}
 \rho_\varphi\simeq V+3\xi H^{2}\varphi^2.
\end{equation}
By comparison with the minimal case as has been studied in Ref.
[20], we see that non-minimal coupling of the scalar field and
gravity on the brane plays an important role on the natural exit
from inflationary phase. On the other hand, it is possible to
achieve the condition $\varepsilon= 1$ before occurrence of
$|\eta|=1$, so that inflation phase terminates when the condition
$\varepsilon= 1$ is fulfilled. This feature is shown in figure $2$
for different values of the non-minimal coupling. We stress that in
this figure the minimal case with $\xi=0$ accounts for natural exit
without additional mechanism too. This is possible in our setup
because of the braneworld effect. In the standard $4$-dimensional
inflationary scenario with one minimally coupled inflaton field, it
is impossible to exit inflationary phase naturally. We note also
that the values of the non-minimal coupling $\xi$ used in the
figures lie within the acceptable range for $\xi$ as has been
obtained in Ref. [25] by confrontation of a non-minimally coupled
phantom cosmology with the recent observations (see also [26]).
\begin{figure}[htp]
\begin{center}\includegraphics{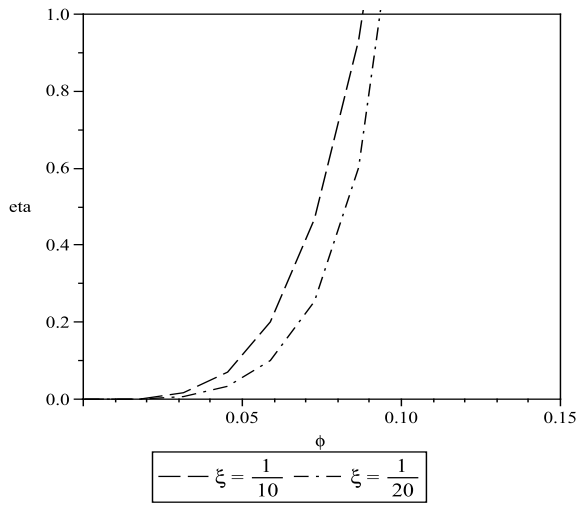} \vspace{6cm}
\end{center}
 \caption{\small {Natural exit from inflation
phase when the condition $|\eta|=1$ is fulfilled before occurrence
of $\epsilon=1$.}}
\end{figure}

\begin{figure}
\begin{center}\includegraphics{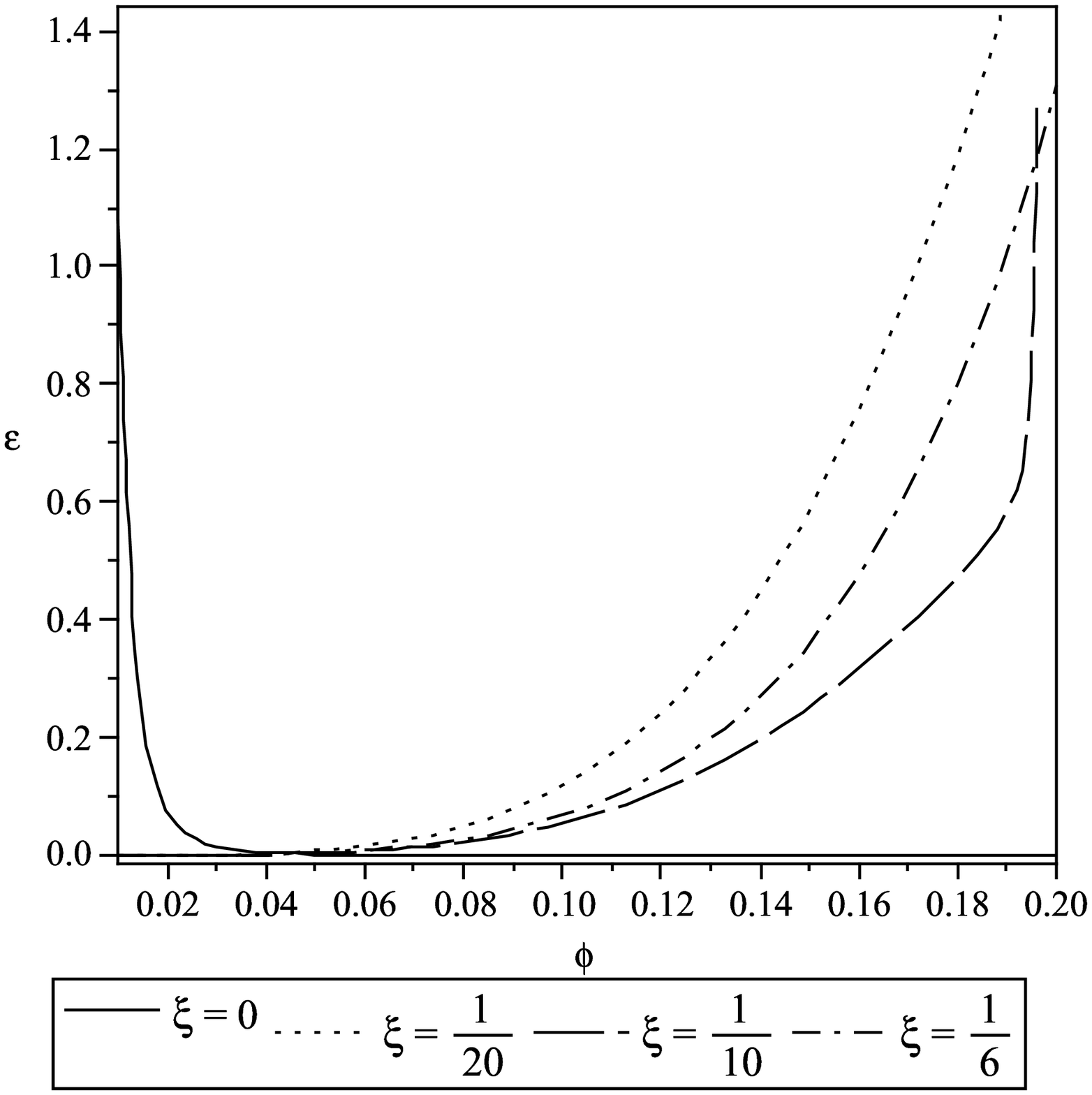} \vspace{5.5cm}\includegraphics{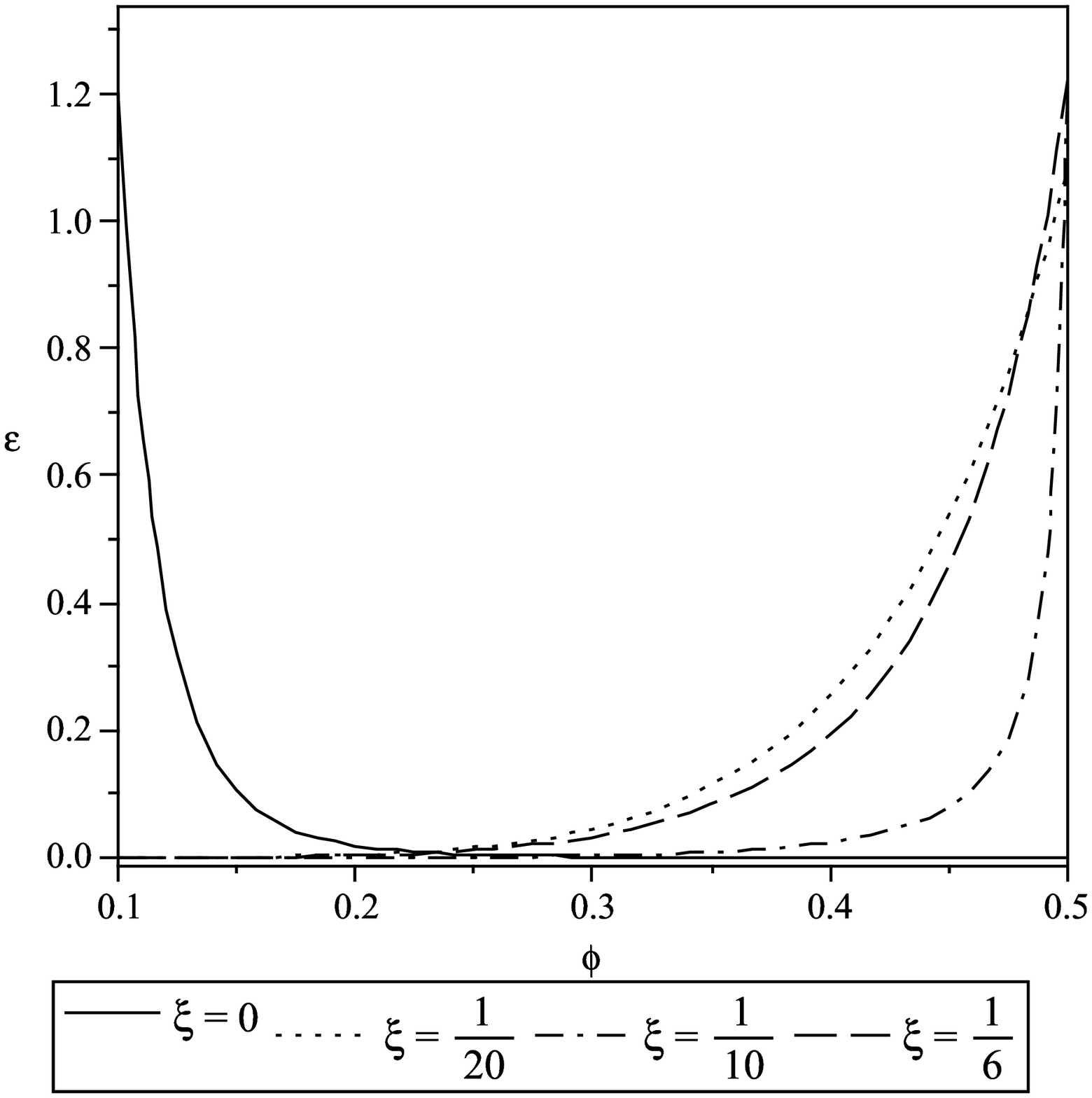}
\end{center}
\caption{\small {Natural exit from inflationary phase
 with various values of the non-minimal coupling for
 $V(\varphi)=V_{0}\varphi^{2}$ (left) and $V(\varphi)=V_{0}\varphi^{4}$ (right)
 in the case that hen the condition $\varepsilon=1$ is fulfilled before occurrence
of $|\eta|=1$. The minimal case is given by $\xi=0$.
 As these figure show, incorporation of the non-minimal coupling causes
 considerable difference in the variation of the slow-roll parameter
 $\varepsilon$, but there is graceful exit as for the minimal case.}}
\end{figure}
Now we focus on the number of e-folds as another important quantity
in a typical inflation scenario. The number of e-folds is defined as
$$N\equiv\int_{t}^{t_{end}}H
dt=\int_{\varphi}^{\varphi_{end}}\frac{H}{\dot{\varphi}}d\varphi$$
In our setup, the number of e-folds at the end of the inflation pase
is given by
\begin{equation}
N=-\int_{\varphi_i}^{\varphi_{e}}\bigg(\frac{3H^2}{\xi
R\varphi+V'}\bigg)d\varphi.
\end{equation}
Using the explicit form of $H^{2}$ with positive sign as given by
equation (13), the number of e-folds in our non-minimal setup takes
the following form
\begin{equation}
N=-\int_{\varphi_{i}}^{\varphi_{e}}\bigg(\frac{3}{\xi
R\varphi+V'}\bigg)\Bigg\{\frac{\lambda}{b^2}\Big[\frac{8\pi}{m_{p}^2}+\frac{8\pi
r_c}{3 m_{p}^2}
  \,b\,(1+\frac{a}{\lambda})\Big]+\frac{2}{b}\,\sqrt{\frac{1}{b^2}+\frac{r_c}{b}(1+\frac{a}{\lambda})}\Bigg\}d\varphi
\end{equation}
with $a$ and $b$ as defined in (14).\,  $\varphi_{i}$  denotes the
value of the scalar field $\varphi$ when universe scale observed
today crosses the Hubble horizon during inflation, while
$\varphi_{e}$ is the value of the scalar field when the universe
exits the inflationary phase. To study the effect of the non-minimal
coupling on the number of e-folds, we define
$$
g\equiv\bigg(\frac{3}{\xi
R\varphi+V'}\bigg)\Bigg\{\frac{\lambda}{b^2}\Big[\frac{8\pi}{m_{p}^2}+\frac{8\pi
r_c}{3 m_{p}^2}
  \,b\,(1+\frac{a}{\lambda})\Big]+\frac{2}{b}\,\sqrt{\frac{1}{b^2}+\frac{r_c}{b}(1+\frac{a}{\lambda})}\Bigg\}
$$ which
is the integrand of equation (21). The number of e-folds is
proportional to the area enclosed between the $g$-curve and
horizontal axis from $\varphi_{i}$ to $\varphi_{e}$. Figure $3$
shows the variation of  $g$ versus $\varphi$ for different values of
the non-minimal coupling and for two different energy scales. As the
value of the non-minimal coupling increases, the enclosed area
between $g$-curve and horizontal axis decreases leading to the
conclusion that the number of e-folds reduces by increasing the
values of the non-minimal coupling. Comparison with the minimal case
shows also that the number of e-folds reduces by inclusion of a
positive non-minimal coupling.
\begin{figure}
\begin{center}\includegraphics{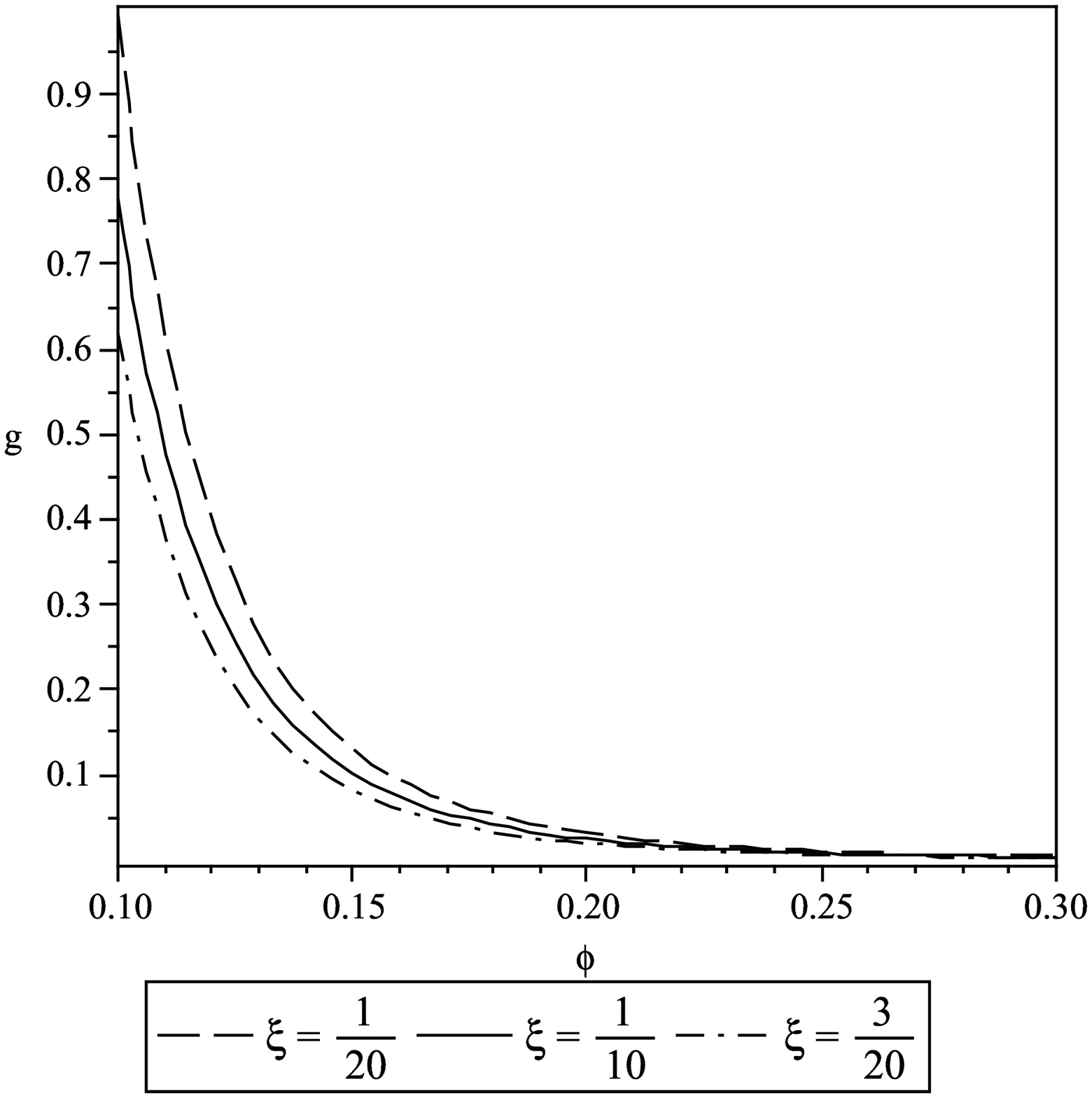} \vspace{5.5cm}\includegraphics{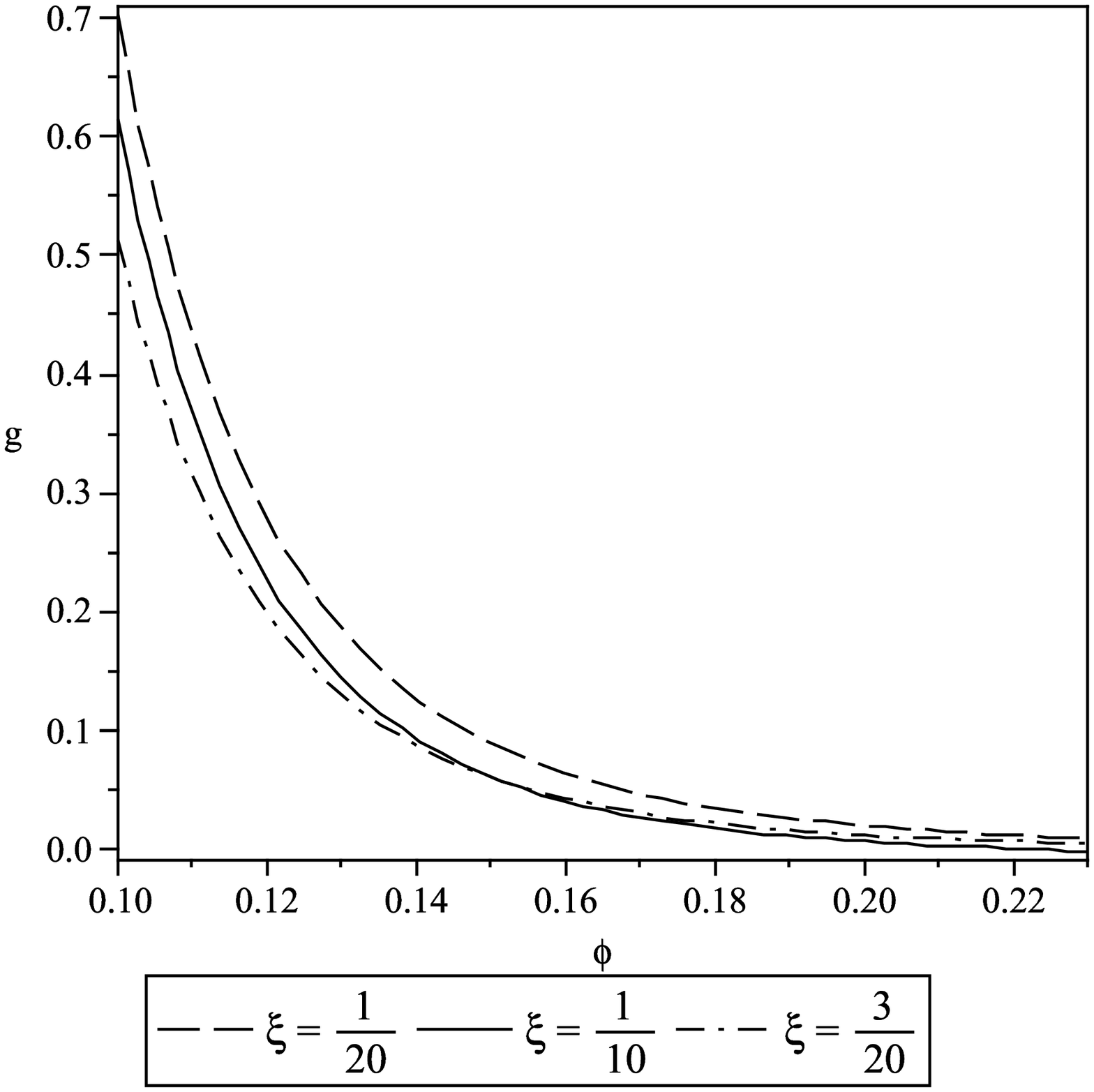}
\end{center}
\caption{\small {$g$ versus $\varphi$ and $\xi$ for two different
regimes: $\lambda\gg 1$ and $\lambda\ll 1$. Number of e-folds is
given by the area bounded between $g$-curve and the horizontal axis
from $\varphi_{i}$ to $\varphi_{e}$ in $g-\varphi$ plot. The number
of e-folds decreases by increasing $\xi$.}}
\end{figure}
Now, we study scalar and tensor perturbations in this non-minimal
model. These perturbations are supposed to be adiabatic on the
brane. We define the scalar curvature perturbation amplitude of a
given mode when re-enters the Hubble radius as follows
\begin{equation}
A_{s}=\frac{2}{5} \frac{H}{\dot{\varphi}}\delta\varphi.
\end{equation}
The scalar spectrum index is defined as follows
\begin{equation}
n_s=1+\frac{d\ln A_{s}^2}{d\ln k}
\end{equation}
The interval in wave number is related to the number of e-folds by
the relation $d \ln k(\varphi)= -d N(\varphi)$. Substituting (3)
into the relation (22) we find for $A_{s}$
\begin{equation}
A_{s}^2=\frac{36}{25}\frac{H^4}{(\xi R\varphi+V')^2} \delta\varphi^2
\end{equation}
Using the slow-roll parameters, we have
\begin{equation}
n_s=1-\frac{7}{2}\varepsilon+ \frac{3}{2}\eta.
\end{equation}
The running of the spectral index which is defined as
\begin{equation}
\alpha_{s}=\frac{dn_{s}}{d\ln k}
\end{equation}
in our model takes the following form
\begin{equation}
\alpha_s=\frac{9}{2}(2\varepsilon\eta-4\varepsilon^2)+
\frac{3}{2}\Big(2\varepsilon\eta-\gamma^2\Big).
\end{equation}

Tensor perturbations are bounded to the brane at long-wavelength and
up to the first order these perturbations are decoupled from matter
on the brane [21]. Therefore, on the large scale we can use the
classical expression for these amplitudes [20]. These amplitudes at
the Hubble crossing are given by
\begin{equation}
A_{T}^{2}=\frac{4}{25\pi}\bigg(\frac{H}{M}\bigg)^{2}\bigg{|}_{k=aH}
\end{equation}
where in our non-minimal case and within the slow-roll approximation
$H$ is given by equation (13)with positive sign, explicitly depended
on the non-minimal coupling, $\xi$. The tensor spectral index is
given by
\begin{equation}
n_{T}\equiv \frac{d\ln A_{T}^{2}}{d\ln k}\approx -2\varepsilon
\end{equation}
where $\varepsilon$ is given by equation (15). The ratio between the
amplitude of tensor and scalar perturbations is given by
\begin{equation}
\frac{A_{T}^{2}}{A_{s}^{2}}\approx \frac{4\pi \Big(\xi R
\varphi+V'(\varphi)\Big)^{2}}{9M^{2} H^{4}}.
\end{equation}
After constructing the basic formalism, we study inflationary
dynamics of this non-minimal model numerically. For this goal, in
which follows and in all of our numerical calculations, we consider
the well-known large-field inflationary potential $V(\varphi)=V_0
\varphi^n$ with $n=2,\,4$ to investigate outcomes of our model. As
has been shown previously in figures $1$ and $2$, this non-minimal
model accounts for natural exit from inflationary phase without any
additional mechanism. In fact, non-minimal coupling of the inflaton
field and gravity on the brane provides a suitable parameter space
for natural exit from inflationary phase. Figure $4$ shows the
variation of $n_{s}$ for different values of the non-minimal
coupling versus the inflaton field.
\begin{figure}[htp]
\begin{center}
\includegraphics{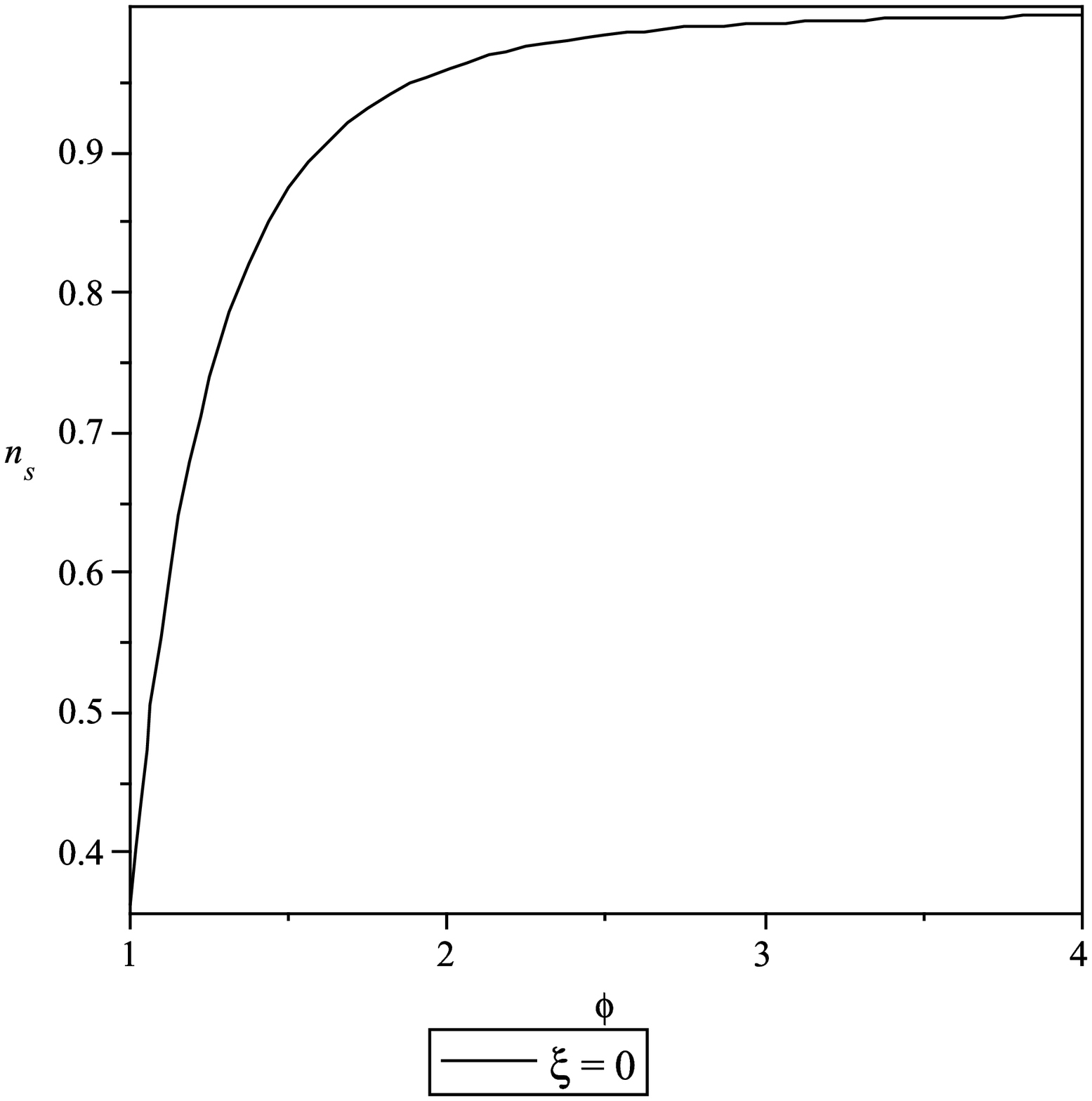} \vspace{5cm}\includegraphics{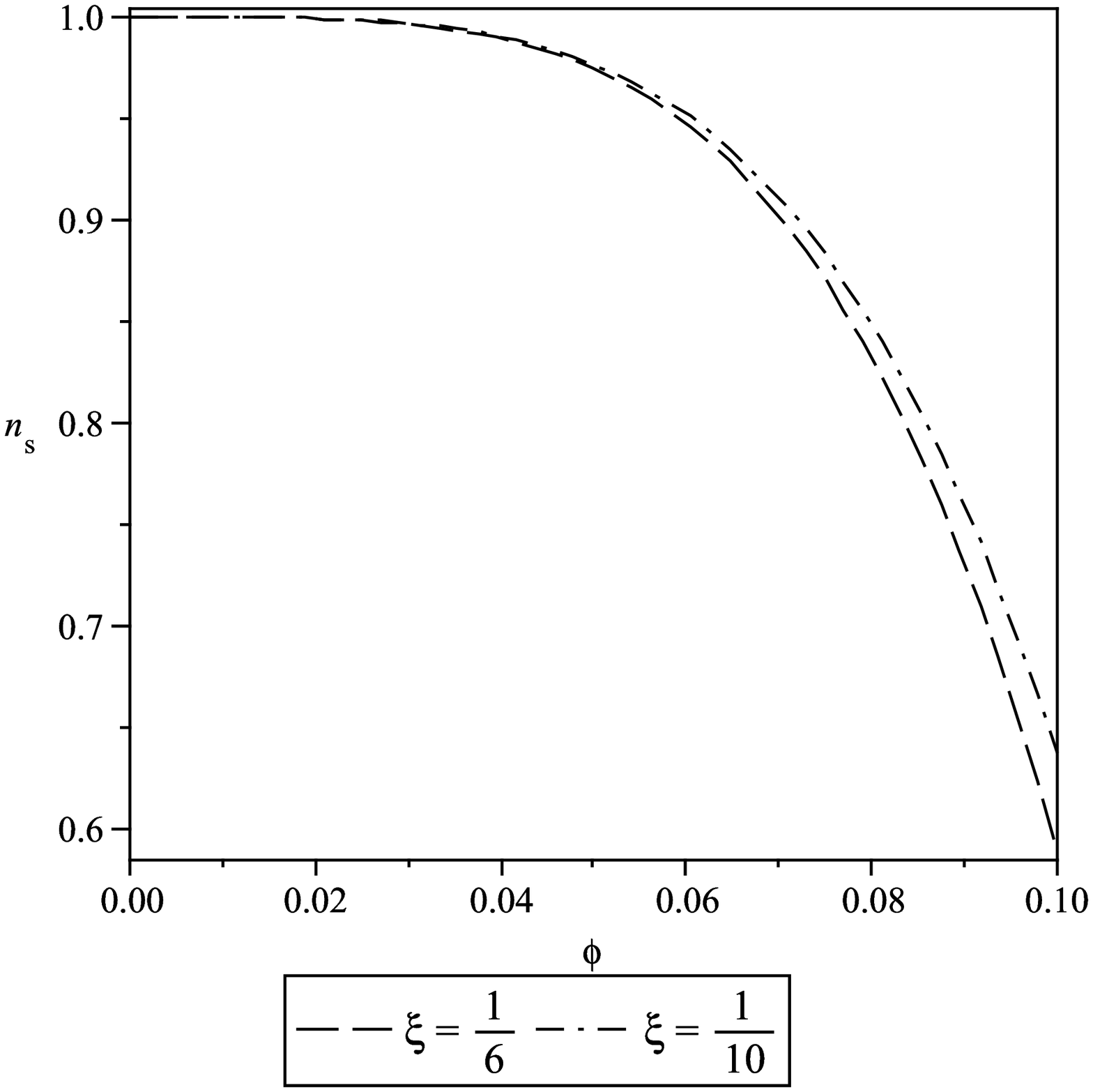}
\end{center}
\vspace{0cm}
 \caption{\small {Variation of $n_{s}$ versus the inflaton field for
 three values of the non-minimal coupling. }}
\end{figure}
Combining WMAP5 with SDSS and SNIa data, the spectral index is given
by ( see for instance [27] and references therein)
\begin{equation}
n_{s}=0.960_{-0.013-0.027}^{+0.014+0.026} \quad\quad (1\sigma,\,
2\sigma\,\, CL).
\end{equation}
This result shows that a red power spectrum is favored and $ n_{s}
> 1$ is disfavored even when gravitational waves are included, which
constrains the models of inflation that can produce significant
gravitational waves, such as chaotic or power-law inflation models,
or a blue spectrum, such as hybrid inflation models. As we have
shown, our non-minimal scenario on the brane gives also a red and
nearly scale invariant power spectrum which excludes blue spectrum.
Since, $0.92\leq n_{s}\leq 1$,\, our numerical calculation shows
that we can constraint $\xi$ to the following range
\begin{equation}
0\leq\xi\leq 0.114\,,
\end{equation}
which lies within the acceptable range for $\xi$ as has been
obtained in Ref. [25] by confrontation of a non-minimally coupled
phantom cosmology with the recent observations. Note that negative
values of $\xi$ corresponding to anti-gravitation are excluded from
our considerations.

If there is running of the spectral index, the constraint on this
running from combined data of WMAP5+SDSS+SNIa is given by
$\frac{dn_{s}}{d\ln k}=-0.032^{+0.021}_{-0.020}$ within the
$1\sigma$ CL [27]. The special case with $n_{s} = 1$ and
$\frac{dn_{s}}{d l\ln k} = 0$ results in the scale invariant
spectrum. The significance of $n_{s}$ and $\frac{dn_{s}}{d l\ln k}$
is that different inflation models motivated by different physics
make specific, testable predictions for the values of these
quantities. Figure $5$ shows the running of the spectral index
versus the inflaton field for different values of the non-minimal
coupling for $V(\varphi)=V_{0}\varphi^{2}$. Evidently, our
non-minimal setup agrees with this dataset and again we can use this
observational data to constraint the values of the non-minimal
coupling. In other words, since $-0.052\leq \alpha_{s} \leq -0.011$,
we can constraint $\xi$ so that $0\leq\xi\leq 0.106$. Comparing this
range of $\xi$ with constraint (32), we see that confrontation of
this non-minimal brane inflation with recent observations leads to
the result that $0\leq \xi \leq 0.106$ which is acceptable from
other studies such as Ref. [25].

\begin{figure}
\begin{center}\includegraphics{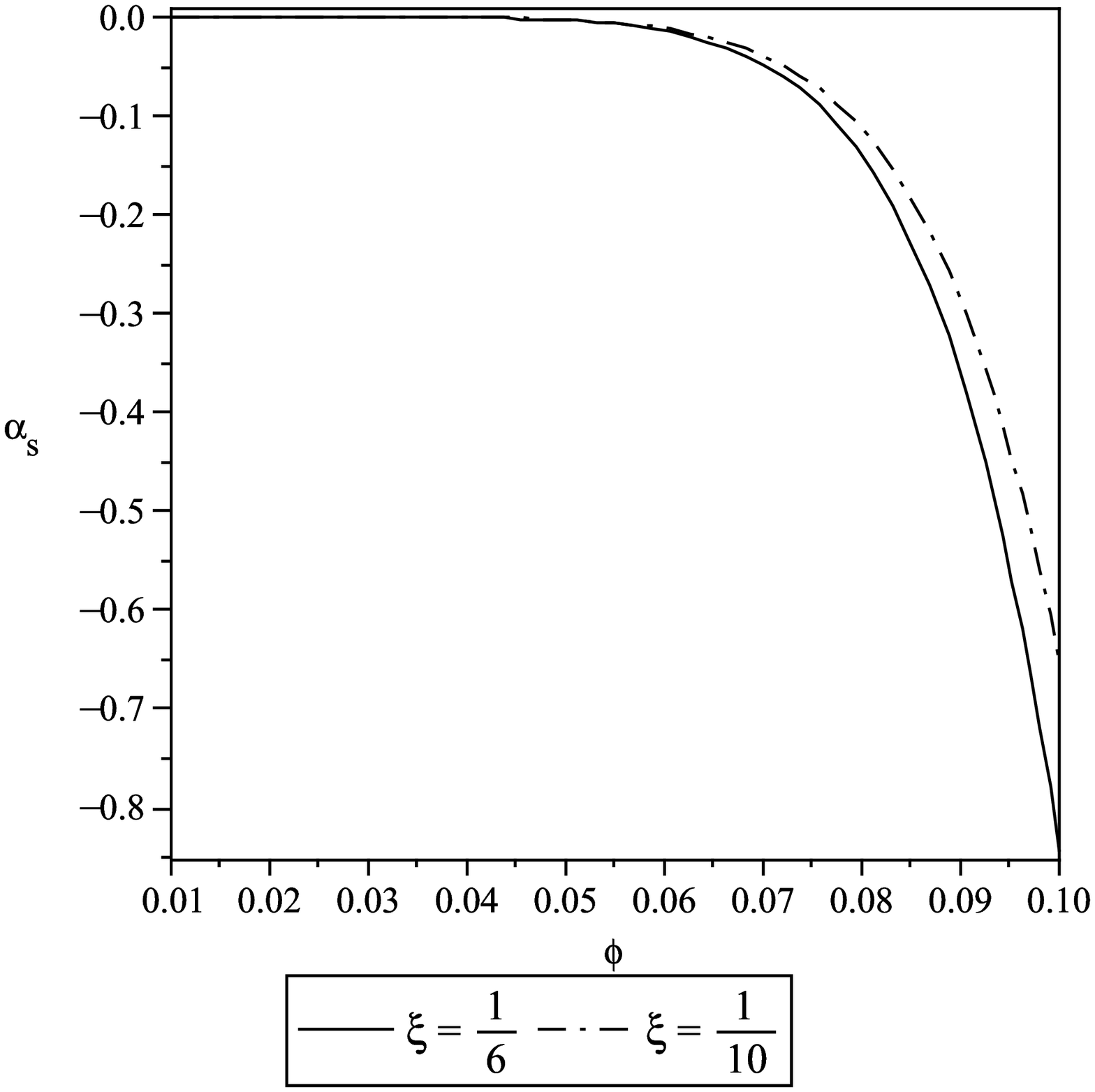} \vspace{5.5cm}\includegraphics{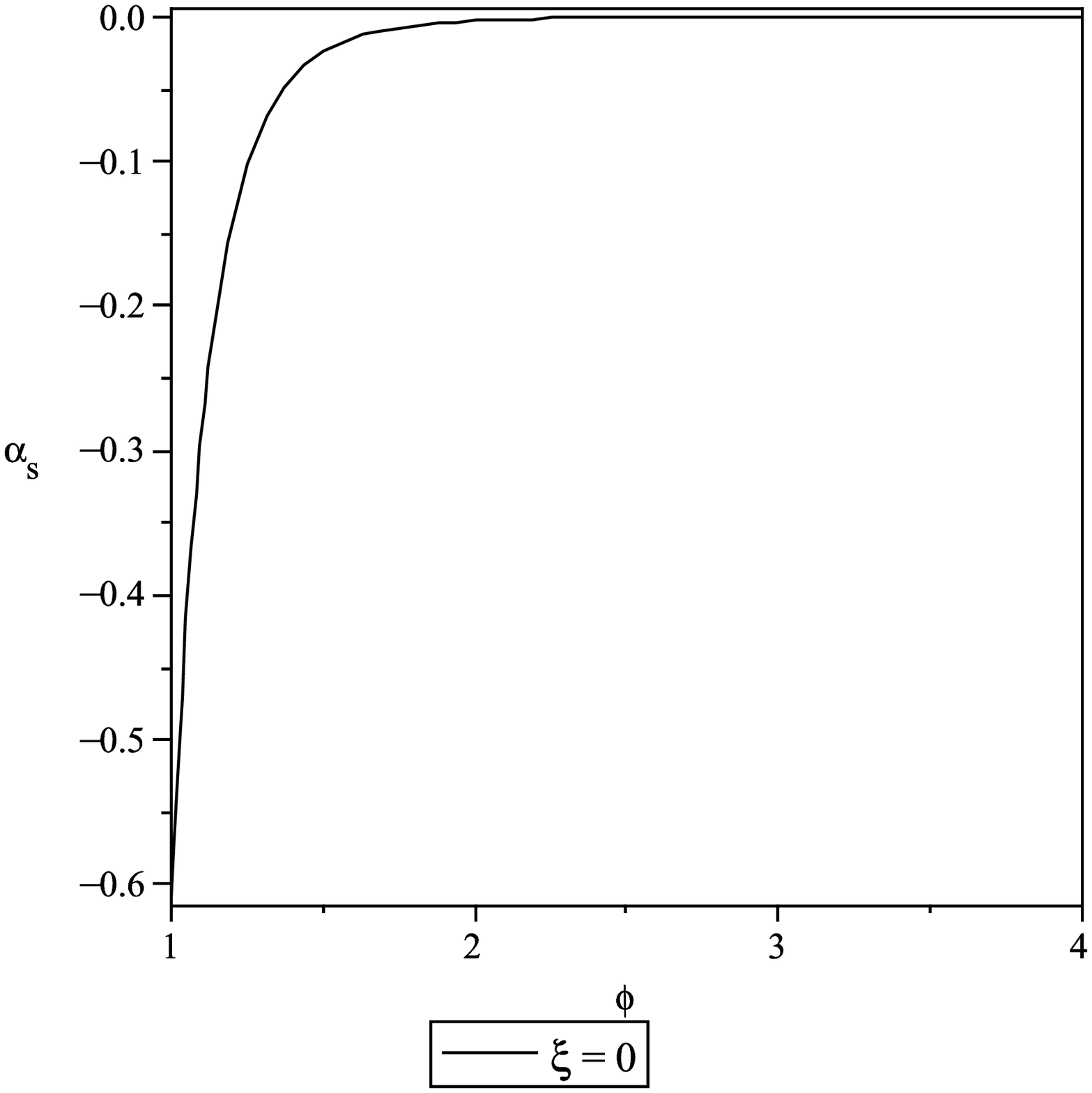}
\end{center}
\caption{\small {$\alpha_{s}$ versus $\varphi$ for different values
of $\xi$.}}
\end{figure}
One important point in our study is the issue of frame. There is a
conformal transformation which transforms the action (1) ( which is
written in the Jordan frame) to corresponding action in the Einstein
frame [5,6]. In the Einstein frame, the gravitational sector is
expressed in terms of a re-scaled scalar field which is minimally
coupled to gravity and evolves in a re-scaled potential, thereby
simplifying the formalism. However, we should keep in mind that the
matter sector is strongly affected by such a conformal
transformation since all of the matter fields are now non-minimally
coupled to the re-scaled metric: in particular, stress tensor
conservation in the matter sector is no longer ensured [28,29]. On
the other hand, as Makino and Sasaki [30] and Fakir {\it et al} [31]
have shown, the amplitude of scalar perturbation in the Jordan frame
exactly coincides with that in the Einstein frame. This proof (for
details see [32]) allows us to calculate the scalar power spectrum
in the Jordan and Einstein frame. As a result, the scalar power
spectrum has no dependence on the choice of frames, i.e., it is
conformally invariant. So, our results can be compared to
observations directly without any ambiguities. This is an important
point since one has to check validity of non-gravity experiments in
Einstein frame. For instance, validity of electro-magnetic related
experiments such as CMB experiment should be checked in Einstein
frame. As Komatsu and Futamase have shown, the scalar power spectrum
is independent on the choice of frames [32].

\section{Summary and Conclusions}

In the inflationary paradigm, inflaton field can interact with other
fields such as Ricci curvature. This interaction is shown by the
non-minimal coupling of the inflaton and gravity in the action of
the scenario. There are many compelling reasons to include an
explicit non-minimal coupling in the action of the theory. So,
naturally we should include non-minimal coupling of gravity and
inflaton field in the action of an inflation scenario. Usually by
incorporation of the non-minimal coupling it is harder to realize
inflation even with potential that are known to be inflationary in
the minimal case. However, inclusion of the non-minimal coupling is
inevitable from field theoretical viewpoint especially their
renormalizability. In this paper we have studied an inflation model
that there is an explicit non-minimal coupling of the inflaton field
and gravity on the Randall-Sundrum II brane. This is an extension of
the study presented in Ref. [20] to scalar-tensor type theories. We
have studied impact of the non-minimal coupling on the dynamics of
this braneworld-inspired inflation. As we have shown, this model
provides a natural exit from the inflationary phase without adopting
any additional mechanism for appropriate values of the conformal
coupling. The number of e-folds decreases by increasing the values
of the conformal coupling, $\xi$. A confrontation with recent
WMAP5+SDSS+ SNIa combined data shows that this model gives a red and
nearly scale invariant power spectrum for
$V(\varphi)=V_{0}\varphi^{2}$ and $V(\varphi)=V_{0}\varphi^{4}$ .
From another viewpoint, comparison of this non-minimal inflation
model with observational data could provide severe constraints on
the values of the non-minimal coupling. In this paper our numerical
analysis shows that $0\leq \xi \leq 0.106$ is favored by non-minimal
inflation on the Randall-Sundrum II brane. Since we have calculated
the first order contributions in slow-roll parameters, our results
are valid in both Jordan and Einstein frames. However, higher order
corrections could lead to different results in these two frames.
Finally, we note that inclusion of the non-minimal coupling of
gravity and inflaton field produces a wider parameter space relative
to minimal case. This wider parameter space provides much more
freedom than single self-interacting scalar field to fit with the
observational data.\\

{\bf Acknowledgments}\\
This work has been supported partially by Research Institute for
Astronomy and Astrophysics of Maragha, Iran.

\end{document}